**UTILIZING BLUETOOTH AND ADAPTIVE SIGNAL CONTROL DATA FOR URBAN ARTERIALS SAFETY ANALYSIS**

**Jinghui Yuan (Corresponding Author), M.S.,** PhD student
Department of Civil, Environmental & Construction Engineering, University of Central Florida
Orlando, FL 32816
Phone: (407) 823-0300 Fax: (407) 823-0300
Email: jinghuiyuan@knights.ucf.edu

**Mohamed Abdel-Aty, Ph.D., P.E.**
Pegasus Professor and Chair
Department of Civil, Environmental & Construction Engineering, University of Central Florida
Orlando, FL 32816
Phone: (407) 823-4535 Fax: (407) 823-3315
Email: M.Aty@ucf.edu

**Ling Wang, Ph.D.**
[1]College of transportation engineering
Tongji University,
1239 Siping Rd, Yangpu Qu, Shanghai Shi, 200000, China
E-mail: LingWang@knights.ucf.edu
[2]Department of Civil, Environmental and Construction Engineering
University of Central Florida，
Orlando, FL 32816，USA

**Jaeyoung Lee, Ph.D.**
Assistant Professor
Department of Civil, Environmental & Construction Engineering, University of Central Florida
Orlando, FL 32816
Phone: (407) 823-4535 Fax: (407) 823-3315
Email: jaeyoung@knights.ucf.edu

**Rongjie Yu, PhD**
Associate Professor
School of Transportation Engineering, Tongji University
4800 Cao'an Road, Jiading District, Shanghai, 201804, China
Email: yurongjie@tongji.edu.cn

**Xuesong Wang, PhD**
Professor
School of Transportation Engineering, Tongji University
4800 Cao'an Road, Jiading District, Shanghai, 201804, China
Email: wangxs@tongji.edu.cn

October 2017

**ABSTRACT**

Real-time safety analysis has become a hot research topic as it can more accurately reveal the relationships between real-time traffic characteristics and crash occurrence, and these results could be applied to improve active traffic management systems and enhance safety performance. Most of the previous studies have been applied to freeways and seldom to arterials. This study attempts to examine the relationship between crash occurrence and real-time traffic and weather characteristics based on four urban arterials in Central Florida. Considering the substantial difference between the interrupted urban arterials and the access controlled freeways, the adaptive signal phasing data was introduced in addition to the traditional traffic data. Bayesian conditional logistic models were developed by incorporating the Bluetooth, adaptive signal control, and weather data, which were extracted for a period of 20 minutes (four 5-minute intervals) before the time of crash occurrence. Model comparison results indicated that the model based on 5-10 minute interval dataset performs the best. It revealed that the average speed, upstream left-turn volume, downstream green ratio, and rainy indicator were found to have significant effects on crash occurrence. Furthermore, both Bayesian random parameters logistic and Bayesian random parameters conditional logistic models were developed to compare with the Bayesian conditional logistic model, and the Bayesian random parameters conditional logistic model was found to have the best model performance in terms of the AUC and DIC values. These results are important in real-time safety applications in the context of Integrated Active Traffic Management.

## INTRODUCTION

Urban arterials play a critical role in the road network system as they provide the high-capacity network for travel within urban areas as well as the access to roadside activities. Meanwhile, urban arterials suffer from serious traffic safety issues. Take Florida as an example, over 51% of crashes have occurred on urban arterials in 2014. Substantial efforts have been made by previous researchers to reveal the relationship between crash frequency on urban arterials and all the possible contributing factors such as roadway geometric, traffic characteristics, etc. (El-Basyouny and Sayed, 2009; Gomes, 2013; Greibe, 2003; Wang et al., 2015b). However, these studies were conducted based on static and highly aggregated data (e.g., Annual Average Daily Traffic (AADT), annual crash frequency).

Recently, an increasing number of studies investigated the crash likelihood on freeways by using real-time traffic and weather data (Abdel-Aty et al., 2004; Abdel-Aty et al., 2012; Ahmed et al., 2012a; Lee et al., 2003; Oh et al., 2001; Xu et al., 2013a; Xu et al., 2013b; Yu and Abdel-Aty, 2014; Yu et al., 2014; Zheng et al., 2010). However, little research has been conducted on the real-time safety analysis of urban arterial (Theofilatos, 2017; Theofilatos et al., 2017), although the real-time traffic and weather data are available on many major arterials. This might be due to the substantial difference in the traffic flow characteristics, data availability, and even crash mechanism between urban arterials and freeways, thus it is inappropriate to simply transfer the same research framework from freeways to urban arterials. More specifically, the interrupted traffic flow on urban arterials is highly influenced by the traffic signals (Cai et al., 2014; Wang et al., 2017b), which is quite different from the uninterrupted flow on freeways. Therefore, the crash risk on urban arterials might be associated with not only real-time traffic flow characteristics but also the real-time signal phasing, which has not been considered in previous research (Theofilatos, 2017; Theofilatos et al., 2017). Moreover, those pioneering studies on the real-time safety analysis of urban arterials were based on one-hour aggregated traffic parameters prior to crash occurrence, which is not really exact "real-time" as the traffic flow are likely to differ within one hour.

In terms of real-time traffic data, most of the previous studies were based on inductive loop detectors (ILDs) (Abdel-Aty et al., 2008; Abdel-Aty et al., 2012; Zheng et al., 2010). ILDs are the most commonly used sensors in traffic management, however, there are some inherent problems with it, such as high failure rates and difficulty with maintenance, especially for arterials. Recently, several studies tried to conduct real-time safety analysis for freeways based on the traffic data collected from nonintrusive detectors, such as automatic vehicle identification system (AVI) (Ahmed et al., 2012a, b; Ahmed and Abdel-Aty, 2012) and remote traffic microwave sensor (RTMS) (Ahmed and Abdel-Aty, 2013; Shi and Abdel-Aty, 2015). AVI is used mainly for toll collection and travel time estimation while RTMS is mostly used for operation and incident management. The speed data collected from different detectors are quite different, AVI and Bluetooth detectors measure space mean speed, whereas RTMS and ILDs measure time mean speed. As to the data availability, AVI and RTMS are usually available on freeways, and the possible available real-time traffic data on urban arterials are ILDs, Bluetooth, and floating car data (FCD). To the knowledge of the authors, there is no real-time safety analysis has been carried out using traffic data from Bluetooth detectors.

Above all, this study aims to investigate the relationship between crash occurrence on urban arterials and real-time traffic, signal phasing, and weather characteristics by utilizing data from multiple sources, i.e., Bluetooth, weather, and adaptive signal control datasets.

## LITERATURE REVIEW
### Aggregated Arterial Safety Analysis
A number of studies have explored the effects of various road geometric design and traffic characteristics on arterial safety based on aggregated data. As to road geometric design, high crash frequency was found to be associated with high intersection density (Bonneson and McCoy, 1997; El-Basyouny and Sayed, 2009; Wang and Yuan, 2017; Wang et al., 2016) and access density (Bonneson and McCoy, 1997; Wang and Yuan, 2017; Wang et al., 2016). The number of lanes was found to be positively correlated with crash occurrence (El-Basyouny and Sayed, 2009; Gomes, 2013; Wang et al., 2015b). In addition, an increased segment length (El-Basyouny and Sayed, 2009; Wang et al., 2015b) and decreased lane width (Yanmaz-Tuzel and Ozbay, 2010) tend to increase the crash frequency.

In terms of traffic related contributing factors, traffic volume and travel speed have been found to be significantly associated with the crash frequency on arterials. Traffic volume (represented by AADT, hourly volume, etc.) has been widely demonstrated to be positively correlated with crash frequency (El-Basyouny and Sayed, 2009; Gomes, 2013; Wang et al., 2015b). While the safety effects of travel speed are not consistent among existing studies, many studies suggested that higher average speed tends to increase the crash frequency (Aarts and Van Schagen, 2006; Elvik, 2009; Nilsson, 2004; Taylor et al., 2002), as higher speed increases the drivers' overall stopping distance which may in turn increase the probability of crash occurrence (Wang et al., 2013). However, some researchers found that the average speed is negatively associated with crash frequency (Baruya, 1998; Stuster, 2004).

Moreover, Pei et al. (Pei et al., 2012) evaluated the relationship between speed and crash risk with respect to distance and time exposure, they found that the correlation between speed and crash risk is positive when distance exposure (i.e., vehicle kilometers travelled) is considered, but negative when time exposure (i.e., vehicle hours travelled derived by multiplying traffic volume by average travel time) is used. Wang et al. (Wang et al., 2015b) utilized the Floating Car Data (FCD) to calculate average speeds during peak and off-peak hours, and then developed crash prediction models for peak and off-peak separately. The model results indicated that average travel speed was not significantly related to crash frequency during the off-peak period, however, during the peak period, a significant positive relationship between average speed and crash frequency was demonstrated. More recently, Imprialou et al. (Imprialou et al., 2016) proposed a new condition-based approach to aggregate the crashes according to the similarity of their pre-crash traffic and geometric conditions, and then compared it with the traditional segment-based aggregation approach. The results showed that average speed was significantly positively associated with crash occurrence in the condition-based model, while the relationship was found to be negative in the segment-based model. In conclusion, the inconsistent findings of the safety effects of travel speed might be caused by the inaccuracy of data aggregation, as the aggregated data cannot represent the actual traffic circumstance when the crashes have occurred. At this point, more disaggregated real-time analysis should be conducted for urban arterials to figure out the underlying relationship between crash occurrence and traffic characteristics.

**Real-time Crash Risk Analysis**

Real-time crash risk analysis has been widely adopted to reveal crash occurrence precursors by investigating the differences in traffic conditions between crash and non-crash events. As crash risk analysis is a typical binary classification problem, the most commonly used methods are the matched case-control logistic models (Abdel-Aty and Pande, 2005; Abdel-Aty et al., 2004; Ahmed and Abdel-Aty, 2012; Xu et al., 2012; Zheng et al., 2010), Bayesian logistical models (Ahmed et al., 2012a; Shi and Abdel-Aty, 2015; Wang et al., 2017a; Wang et al., 2015a; Yu et al., 2014), Bayesian random effect logistic models (Shi and Abdel-Aty, 2015; Yu et al., 2016), Bayesian random parameter logistic models (Shi and Abdel-Aty, 2015; Xu et al., 2014; Yu and Abdel-Aty, 2014; Yu et al., 2017). Besides, several approaches of data mining such as neural networks (Abdel-Aty and Pande, 2005; Abdel-Aty et al., 2008), support vector machines (Yu and Abdel-Aty, 2013; Yu and Abdel-Aty, 2014), and Bayesian networks (Hossain and Muromachi, 2012; Sun and Sun, 2015) were also applied to evaluate the real-time crash risk.

In order to identify the crash-prone conditions, huge efforts have been made to investigate the relationship between real-time crash risk and various traffic parameters and weather-related variables. Generally, the average speed was found to be negatively correlated with crash likelihood (Abdel-Aty et al., 2012; Ahmed et al., 2012a, b; Ahmed and Abdel-Aty, 2012; Shi and Abdel-Aty, 2015; Xu et al., 2012; Yu et al., 2016). The speed variation in the form of speed standard deviation or coefficient of speed variation was found to have significant positive effects on crash occurrence (Abdel-Aty et al., 2004; Abdel-Aty et al., 2012; Ahmed et al., 2012a, b; Ahmed and Abdel-Aty, 2012; Xu et al., 2012; Zheng et al., 2010). Intuitively, higher traffic volume contributes to higher crash risk (Roshandel et al., 2015). Moreover, several studies (Hossain and Muromachi, 2012; Shi and Abdel-Aty, 2015) reported that the congestion index is positively correlated with crash occurrence. With respect to weather related variables, adverse weather is usually associated with increased crash risk (Ahmed et al., 2012a; Xu et al., 2013a).

In summary, all the above real-time safety analyses were focused on the freeways, while urban arterials have seldom been analyzed. Theofilatos (2017) was the first to investigate crash likelihood and severity by exploiting real-time traffic and weather data collected from urban arterials. He found that both the variation in occupancy and logarithm of the coefficient of variation of flow are positively associated with crash occurrence. However, the traffic parameters were aggregated to 1-hour interval, which might be too large to capture the short-term traffic status prior to crash occurrence. Moreover, it is worth noting that the crash risk of urban arterials might be highly influenced by signal operation, while it has never been examined in real-time safety analysis.

## DATA PREPARATION

The roads chosen are four urban arterials in Orlando, Florida, as shown in Figure 1. Initially, 72 road segments in both directions were considered in this study, the road segment here mentioned is defined as the segment between adjacent intersections. A total of four datasets were used: (1) crash data from March, 2017 to December, 2017 provided by Signal Four Analytics (S4A); (2) travel speed data collected by 23 IterisVelocity Bluetooth detectors installed at 23 intersections; (3) signal phasing and 15-minute interval traffic volume provided by 23 adaptive signal controllers; (4) weather characteristics collected by the nearest airport weather station.

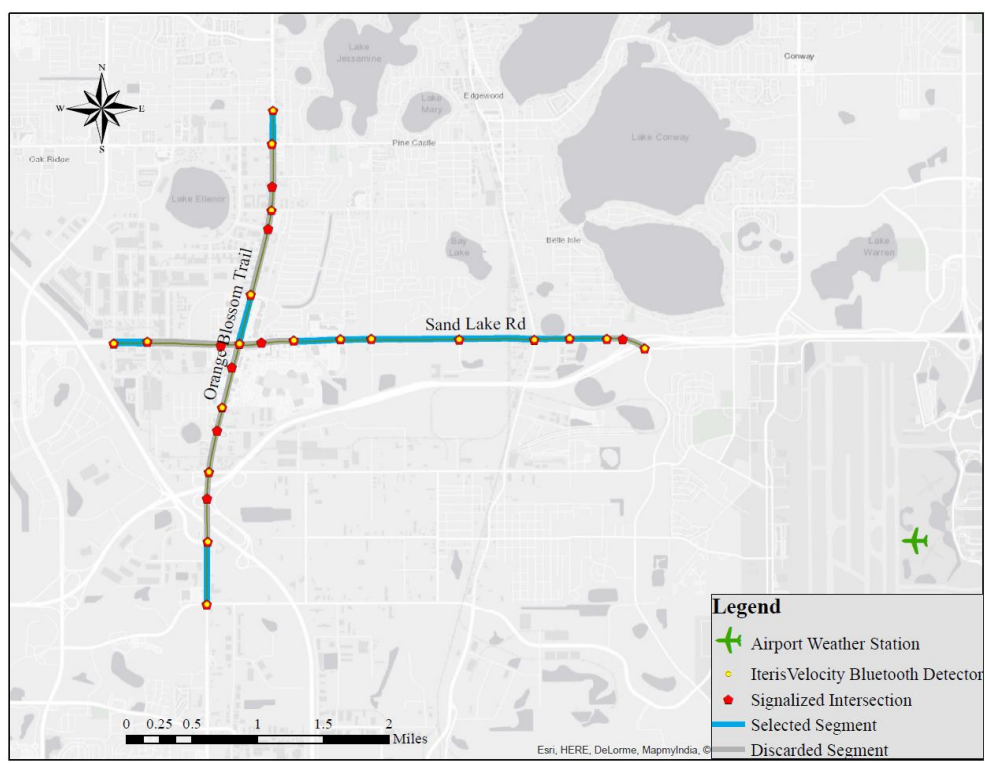

(a) Sand Lake Road and Orange Blossom Trail

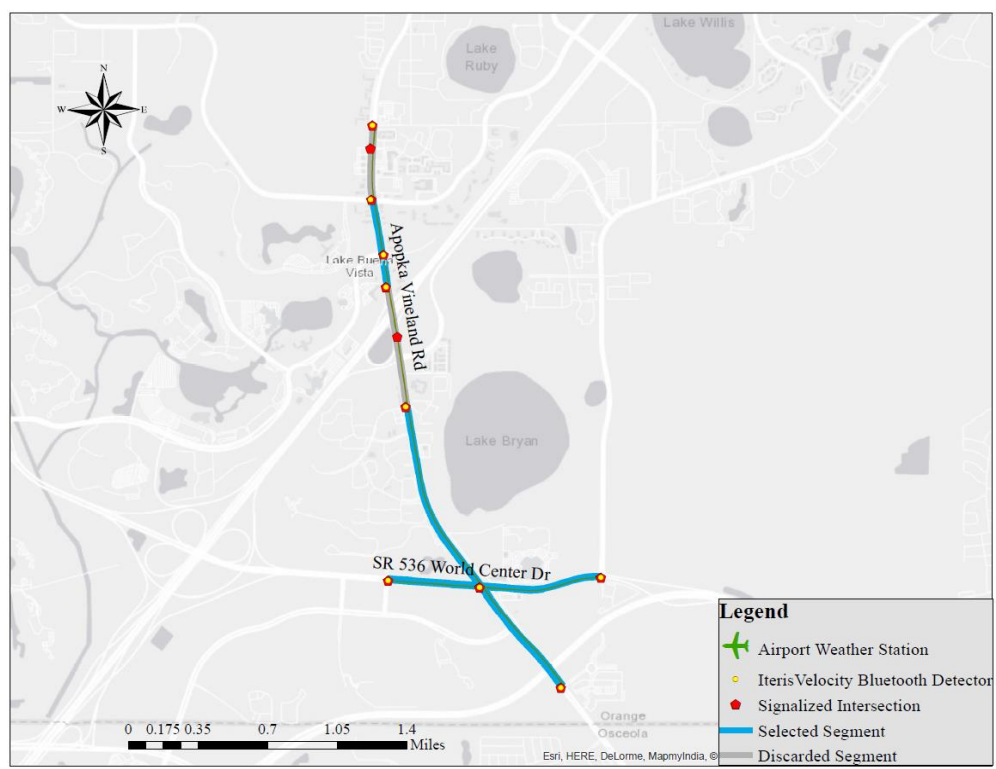

(b) Apopka Vineland Road and SR 536 World Center Drive

**FIGURE 1. Selected Four Urban Arterials**

S4A provides detailed crash information, including crash time, coordinates, severity, type, weather condition, etc. In terms of the crash time information, there are three kinds of time information for each crash, i.e. time of crash occurrence, time reported, and time dispatched. Only the time of crash occurrence was utilized in this study, and the difference between this crash time and the actual crash time is supposed to be within 5 minutes since there exist several efficient and accurate technologies for the police officer to identify the accurate time of crash occurrence, e.g. closed-circuit television cameras and mobile phones.

First, all crashes occurred on the selected arterials from March, 2017 to December, 2017 were collected. After that, based on the attributes of "Type of Intersection" and "First Harmful Event Relation to Junction", all the intersection and intersection-related crashes were excluded. Meanwhile, all the crashes that occurred under the influence of alcohol and drugs were excluded. After these filtering processes, a total of 523 crashes remained and these crashes were assigned to the corresponding road segments.

Matched case-control design was employed in this study to explore the effects of traffic, signal, and weather related variables while eliminating the effects of other confounding factors through the design of study. First, all the crash events were collected, and for each selected crash, several confounding factors, i.e., segment ID, time of day, and day of the week, were selected as matching factors. Therefore, a group of non-crash events could be identified by using these matching factors and then a specific number of non-crash events could be randomly selected from this group of non-crash events for every crash (FIGURE 2). The number of non-crash events m corresponding to a crash event is preferred to be fixed in the entire analysis. As stated in Hosmer Jr et al. (2013), the value of m was commonly chosen from one to five. In addition, Abdel-Aty et al. (2004) found that there is no significant difference when m changing from one to five. Therefore, the control-to-case ratio of 4:1 was adopted in this study, which is consistent with previous research (Abdel-Aty et al., 2008; Ahmed and Abdel-Aty, 2013; Ahmed et al., 2012b; Ahmed and Abdel-Aty, 2012; Shi and Abdel-Aty, 2015; Xu et al., 2012; Yu et al., 2016; Zheng et al., 2010). Consequently, 4 non-crash events from the same road segment, time of day, and day of week were extracted for each crash event. Besides, these non-crash events were extracted only when there is no crashes occurring within 3 hours before or after the non-crash event on the same road segment.

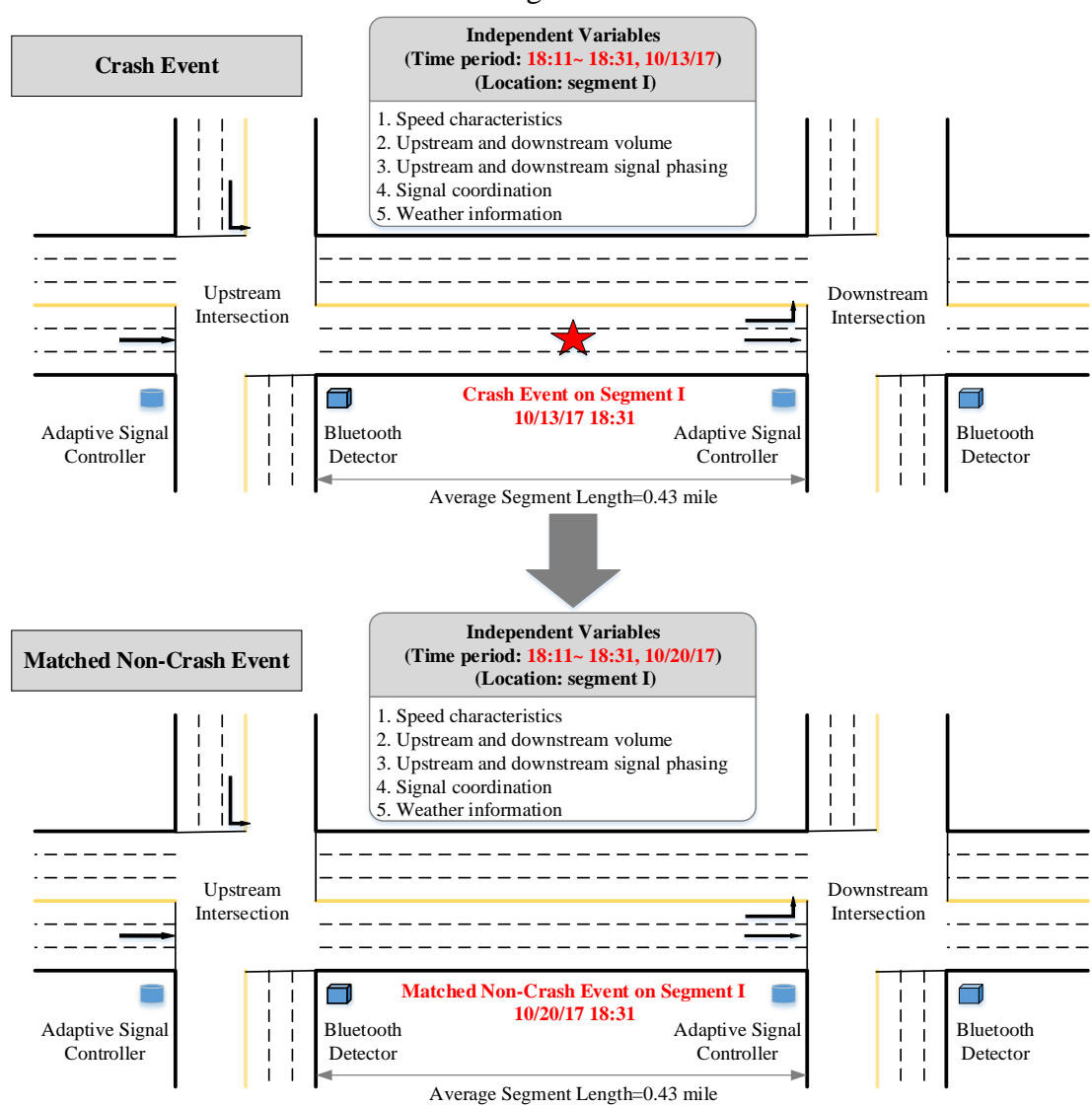

**FIGURE 2. Illustration of Matched Case-Control Design**

Bluetooth data provides the travel time and space-mean speed of the detected vehicle for each segment. Bluetooth detectors can only detect the vehicles equipped with Bluetooth device and the device is working at discoverable mode. The space-mean speed of each vehicle on a specific segment is calculated as the segment length divided by the travel time of each detected vehicle on the segment based on the detection data of two Bluetooth detectors located at the two contiguous intersections. The procedure of Bluetooth data collection is illustrated in FIGURE 3. In order to mitigate the impact of signal delay, the vehicle-level travel speed data were filtered by the algorithm which only keeps the data sample within 75% of the interquartile range of the preceding 15 samples on the same segment, this filtering algorithm could filter out those biased samples which might be highly influenced by the signal delay.

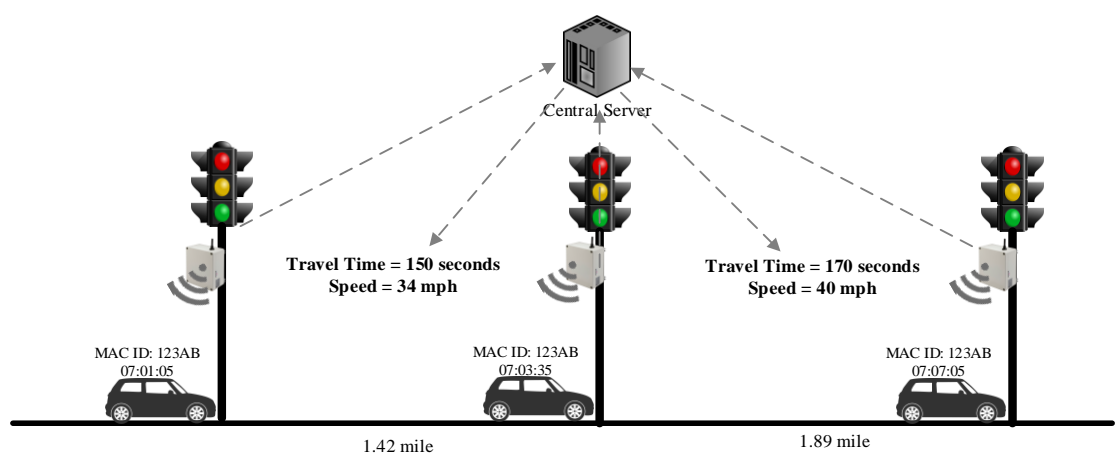

**FIGURE 3. Illustration of Bluetooth Data Collection**

If there is no Bluetooth detector on one of the contiguous intersection (FIGURE 4), the travel speed on the segment will be decreased after including the intersection delay, thus, all the segments with missing Bluetooth detector on either contiguous intersections were deleted. Consequently, only 32 road segments were selected for data collection (FIGURE 1).

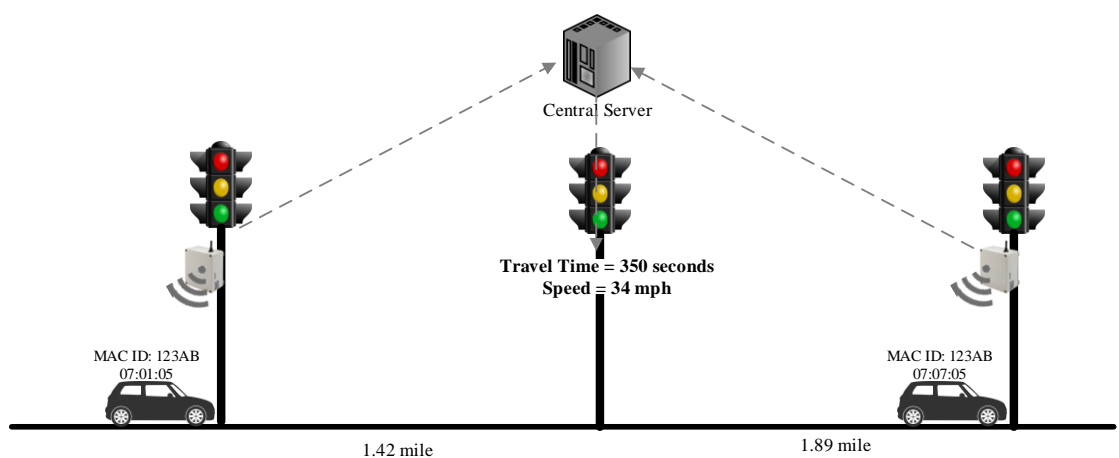

**FIGURE 4**. **Illustration of Excluded Bluetooth Segment**

It is worth noting that the Bluetooth overall sampling rate is 6.05%, which is higher than the threshold suggested by the previous studies (Chen and Chien, 2000; Long Cheu et al., 2002), which stated that a floating car sample of just 3% of the vehicle population is sufficient for a 95% confidence level in travel time and speed estimates. The real-time travel speed data were extracted for a period of 20 minutes (divided into four 5-minute time slices) prior to crash occurrence. For example, if a crash occurred on segment-15 at 15:00, the corresponding travel speed data from 14:40 to 15:00 were extracted and named as time-slices 1, 2, 3, and 4. The distribution histogram of the 5-minute Bluetooth sample frequency is shown in FIGURE 5, if the number of vehicles that are detected within any time slice is lower than 2 (17.12%), then the corresponding crashes were excluded. Finally, a total of 273 crashes were used in the analysis.

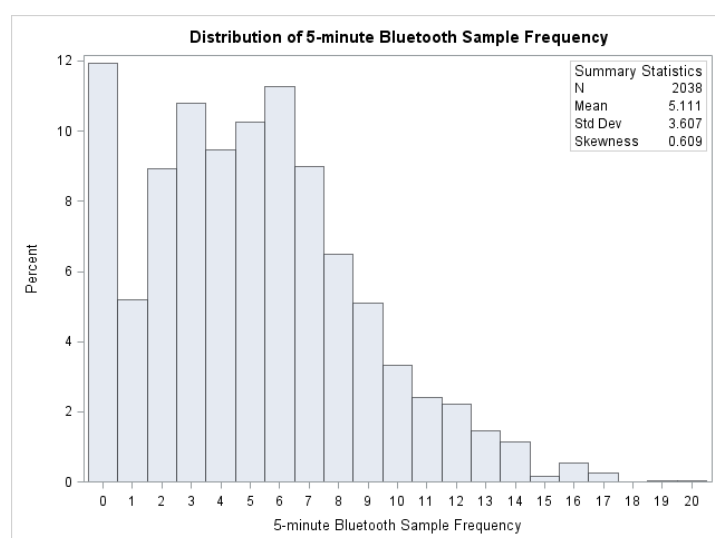

**FIGURE 5. Distribution of 5-minutes Bluetooth Sample Frequency**

The adaptive signal control system at a signalized intersection is operated based on the video detectors installed on the approaches, which can detect the real-time queue length, maximum waiting time, and traffic volume by movement. This system archives the real-time signal phasing, queue length, waiting time, and 15-minute aggregated traffic volume data. Since the right-turn vehicles are unprotected at the intersection, the traffic volume data only include the through and left-turn vehicles. As shown in FIGURE 2, the upstream volume of the segment consists of the through and left-turn traffic volume coming from the upstream intersection, while the downstream volume of the segment consists of the through and left-turn traffic volume approaching into the downstream intersection. Since the archived volume data are aggregated by 15 minutes, therefore, the traffic volume during 5-mintue interval was proportionally calculated based on the assumption that the traffic volume within 15-minute interval are evenly distributed.

The 5-minute through green ratio for the contiguous upstream and downstream intersections were collected for the period of 4 time slices prior to the reported crash time. Also, the 5-minutes signal coordination between the contiguous upstream and downstream intersections was collected. As shown in FIGURE 6, the signal coordination is the total maximum bandwidth ("windows" of green for traveling platoons) between the upstream and downstream signals during the periods of 4 time slices prior to the reported crash time. The ideal offset, which is calculated by the segment length divided by the corresponding speed limit, was adopted to represent the offset between the upstream and downstream intersections.

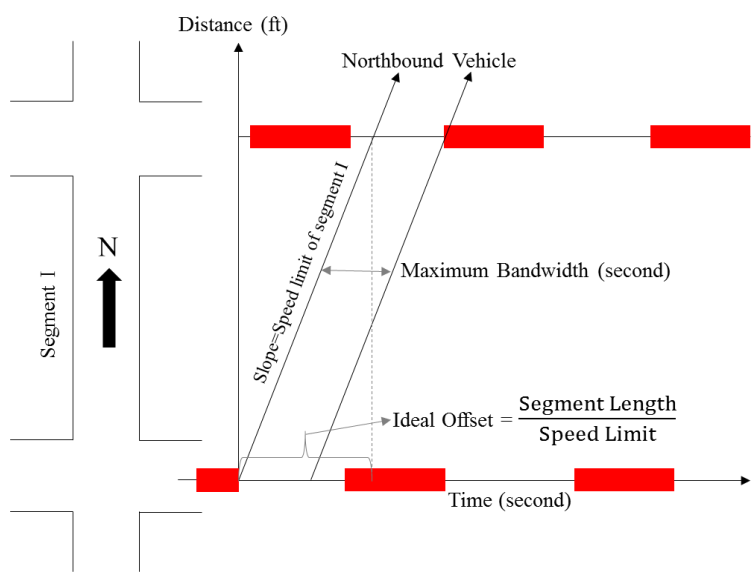

**FIGURE 6. Illustration of maximum bandwidth and signal coordination**

Two weather related variables (rainy weather indicator and visibility) were collected from the nearest airport weather station, which is located at the Orlando international airport (as shown in FIGURE 1). Since the weather data is not recorded continuously, once the weather condition changes and reaches a preset threshold, a new record will be added to the archived data. Therefore, for each specific crash, based on the reported crash time, the closest weather record prior to the crash time has been extracted and used as the crash time weather condition, which is identical for four time slices.

In order to validate the weather data collected by the airport weather station with the weather condition reported in the crash report. The weather type information of each crash event collected from two data sources was selected to conduct a cross table analysis. The weather type information reported in the crash report including clear (76.44%), cloudy (13.29%), and rain (10.27%), which were converted into a binary variable (rainy and normal) to compare with the rainy weather indicator collected by the airport weather station. The results indicated that the accuracy ((True positive + True negative)/Total sample size) of weather station is 92%.

The final dataset includes 1365 observations (273 crash events and 1092 non-crash events), which were then divided into training (80%: 218 crash events) and validation (20%:

55 crash events) datasets. The summary statistics of the final dataset for all the traffic, signal, and weather-related variables are as shown in Table 1.

**TABLE 1. Summary of Variables Descriptive Statistics (Crash and Non-crash Events)**

| Variables | Time Slice | Description | Mean | Std dev. | Min | Max |
|---|---|---|---|---|---|---|
| Crash_count | - | Number of crashes for each segment | 9.10 | 7.50 | 1.00 | 29.00 |
| Avg_Speed | 1 | Average speed within 5-minute interval (mph) | 25.91 | 10.18 | 4.88 | 55.00 |
| | 2 | | 26.07 | 10.01 | 4.00 | 56.00 |
| | 3 | | 26.40 | 10.05 | 4.00 | 58.00 |
| | 4 | | 26.21 | 9.84 | 4.33 | 59.33 |
| Std_Speed | 1 | Speed standard deviation within 5-minute interval (mph) | 9.86 | 5.20 | 0.00 | 30.41 |
| | 2 | | 10.06 | 5.02 | 0.00 | 31.01 |
| | 3 | | 10.00 | 5.12 | 0.00 | 36.77 |
| | 4 | | 10.11 | 5.22 | 0.00 | 28.54 |
| Up_Vol | 1 | Number of vehicles coming from the upstream intersection within 5-minute interval | 108.85 | 53.55 | 0.00 | 346.67 |
| | 2 | | 109.00 | 53.81 | 0.00 | 346.67 |
| | 3 | | 108.25 | 53.04 | 0.00 | 316.67 |
| | 4 | | 107.87 | 54.54 | 0.00 | 491.33 |
| Down_Vol | 1 | Number of vehicles approaching into the downstream intersection within 5-minute interval | 123.28 | 56.81 | 0.00 | 869.33 |
| | 2 | | 123.38 | 56.05 | 0.00 | 869.33 |
| | 3 | | 122.85 | 56.54 | 0.00 | 869.33 |
| | 4 | | 122.96 | 55.76 | 0.00 | 557.33 |
| Up_Vol_LT | 1 | Number of left turn vehicles coming from the upstream intersecting road segment within 5-minute interval (Figure 2) | 10.19 | 18.15 | 0.00 | 146.67 |
| | 2 | | 10.14 | 18.00 | 0.00 | 134.93 |
| | 3 | | 10.18 | 18.09 | 0.00 | 142.67 |
| | 4 | | 10.11 | 17.80 | 0.00 | 142.67 |
| Down_Vol_LT | 1 | Number of left turn vehicles approaching into the downstream intersection within 5-minute interval (Figure 2) | 16.12 | 14.76 | 0.00 | 118.33 |
| | 2 | | 16.09 | 15.25 | 0.00 | 149.67 |
| | 3 | | 16.14 | 15.59 | 0.00 | 149.67 |
| | 4 | | 16.14 | 15.79 | 0.00 | 149.67 |
| Up_Green_Ratio | 1 | The percentage of green time for through vehicle in the upstream intersection within 5-minute interval (%) | 47.87 | 18.32 | 4.00 | 100.00 |
| | 2 | | 46.96 | 17.57 | 12.67 | 94.67 |
| | 3 | | 47.11 | 18.14 | 8.33 | 100.00 |
| | 4 | | 47.78 | 17.96 | 10.67 | 100.00 |
| Down_Green_ratio | 1 | The percentage of green time for through vehicle in the downstream intersection within 5-minute interval (%) | 46.97 | 18.66 | 9.00 | 100.00 |
| | 2 | | 47.11 | 18.76 | 7.67 | 93.67 |
| | 3 | | 46.17 | 17.99 | 6.67 | 100.00 |
| | 4 | | 46.76 | 18.85 | 7.67 | 92.33 |
| Signal_Coordination | 1 | Total bandwidth divided by the upstream green time within 5-minute interval | 0.66 | 0.29 | 0.00 | 1.00 |
| | 2 | | 0.65 | 0.29 | 0.00 | 1.00 |
| | 3 | | 0.65 | 0.29 | 0.00 | 1.00 |
| | 4 | | 0.65 | 0.29 | 0.00 | 1.00 |
| Rainy | - | Binary variable for rainy weather indicator (0 for normal and 1 for rainy) | 0.05 | 0.21 | 0.00 | 1.00 |
| Visibility | - | Visibility (mile) | 9.79 | 1.09 | 1.00 | 10.00 |

## METHODOLOGY

As crash risk analysis is a typical binary classification problem (crash and non-crash), logistic regression model would be the most basic and preferable method. However, since the matched case control design was employed in this study to select the non-crash events rather than the random sample method, which means that the selected non-crash events and the corresponding crash event are within the same stratum. Therefore, conditional logistic regression, which is also known as matched-case control regression, should be more appropriate for this study, which is in line with previous research (Abdel-Aty and Pande, 2005; Abdel-Aty et al., 2004; Ahmed and Abdel-Aty, 2012; Xu et al., 2012; Zheng et al., 2010). In this study, four Bayesian conditional logistic models were developed for the four time slices separately.

Furthermore, many previous research found that random parameters model performs much better than fixed parameters model (Shi and Abdel-Aty, 2015; Xu et al., 2014; Yu and Abdel-Aty, 2014; Yu et al., 2017). Therefore, Bayesian random parameters logistic model and Bayesian random parameters conditional logistic model were also employed based on the best time slice dataset to compare with the Bayesian conditional logistic model. Bayesian approach, which treats the parameters as random variable and incorporates prior knowledge to estimate the posterior distribution of parameters, was adopted in this study. It was claimed that the Bayesian approach provided better fit and reduced uncertainty for parameter estimations than the frequentist approach (Ahmed et al., 2012b).

### Bayesian Conditional Logistic Model

Suppose that there are $N$ strata with 1 crash ($y_{ij}=1$) and $m$ non-crashes ($y_{ij}=0$) in stratum $i$, $i=1$, 2, …, $N$ and $j=0,1,2,…,m$. Let $p_{ij}$ be the probability that the $j$th observation in the $i$th stratum is a crash. This crash probability could be expressed as:

$$y_{ij} \sim Bernoulli(p_{ij}) \tag{1}$$

$$logit(p_{ij}) = \alpha_i + \beta_1 X_{1ij} + \beta_2 X_{2ij} + \cdots + \beta_k X_{Kij} \tag{2}$$

Where $\alpha_i$ denotes the effects of matching variables on crash likelihood for ith stratum; $\boldsymbol{\beta} = (\beta_1, \beta_2, …, \beta_K)$ is the vector of regression coefficients for K independent variables, and all the $\boldsymbol{\beta}$ coefficients are set up with non-informative priors as following normal distributions (0, 1E-6); $\boldsymbol{X_{ij}} = (X_{1ij}, X_{2ij}, …, X_{Kij})$ is the vector of K independent variables.

In order to take the stratification in the analysis of the observed data, the stratum-specific intercept $\alpha_i$ is considered to be nuisance parameters. Suppose the observation $y_{i0}$ is a crash, and $y_{ij}, j = 1, 2, …, m$ are non-crashes, then the conditional likelihood for the ith stratum would be expressed as (Hosmer Jr et al., 2013):

$$l_i(\boldsymbol{\beta}) = \frac{\exp(\sum_{k=1}^{K} \beta_k X_{ki0})}{\sum_{j=0}^{m} \exp(\sum_{k=1}^{K} \beta_k X_{kij})} \tag{3}$$

And the full conditional likelihood is the product of the $l_i(\beta)$ over N strata,

$$L(\boldsymbol{\beta}) = \prod_{i=1}^{N} l_i(\boldsymbol{\beta}) \tag{4}$$

Since the full conditional likelihood is independent of stratum-specific intercept $\alpha_i$, thus Equation 2 cannot be used to estimate the crash probabilities. However, the estimated $\boldsymbol{\beta}$ coefficients are the log-odd ratios of corresponding variables and can be used to approximate the relative risk of an event. Furthermore, the log-odds ratios can also be used to develop a prediction model under this matched case-control analysis. Suppose two observation vectors $\boldsymbol{X_{i1}} = (X_{1i1}, X_{2i1}, \dots, X_{Ki1})$ and $\boldsymbol{X_{i2}} = (X_{1i2}, X_{2i2}, \dots, X_{Ki2})$ from the ith stratum, the odds ratio of crash occurrence caused by observation vector $\boldsymbol{X_{i1}}$ relative to observation vector $\boldsymbol{X_{i2}}$ could be calculated as:

$$\frac{p_{i1}/(1 - p_{i1})}{p_{i2}/(1 - p_{i2})} = \exp[\sum_{k=1}^{K} \beta_k (X_{ki1} - X_{ki2})] \tag{5}$$

The right hand side of Eq. (5) is independent of $\alpha_i$ and can be calculated using the estimated $\boldsymbol{\beta}$ coefficients. Thus, the above relative odds ratio could be utilized for predicting crash occurrences by replacing $\boldsymbol{X_{i2}}$ with the vector of the independent variables in the ith stratum of non-crash events. One may use simple average of each variable for all non-crash observations within the stratum. Let $\overline{\boldsymbol{X}}_{\boldsymbol{i}} = (\overline{X}_{1i}, \overline{X}_{2i}, \dots, \overline{X}_{Ki})$ denote the vector of mean values of non-crash events of the k variables within the ith stratum. Then the odds ratio of a crash relative to the non-crash events in the ith stratum could be approximated by:

$$\frac{p_{i1}/(1 - p_{i1})}{p_i/(1 - p_i)} = \exp[\sum_{k=1}^{K} \beta_k (X_{ki1} - \overline{X}_{ki})] \tag{6}$$

**Bayesian Random Parameters Logistic Model**

Suppose the crash occurrence has the outcomes $y_i$=1 (crash event) and $y_i$=0 (non-crash event) with respective probability $p_i$ and 1-$p_i$, $i$=1, 2,…, $N(m+1)$. *N* and *m* represent the number of strata and the number of control events within each stratum, separately. *N(m+1)* indicates the total number of observations. The random parameters logistic regression can be expressed as follows:

$$y_i \sim Bernoulli(p_i) \tag{7}$$

$$logit(p_i) = \beta_{0i} + \beta_{1i}X_{1i} + \beta_{2i}X_{2i} + \cdots + \beta_{Ki}X_{Ki} \tag{8}$$

$$\beta_{ki} = \beta_k + \varphi_{ki}, \qquad k = 0,1,2, \dots, K \tag{9}$$

$$\varphi_{ki} \sim N(0, \sigma_k^2) \tag{10}$$

Where $\beta_{0i}$ is the random intercept for the ith observation; $\boldsymbol{\beta_i} = (\beta_{1i}, \beta_{2i}, \dots, \beta_{Ki})$ is the vector of K random coefficients for the ith observation; $\boldsymbol{X_{ij}} = (X_{1ij}, X_{2ij}, \dots, X_{Kij})$ is the vector of K independent variables for the ith observation; $\varphi_{ki}$ is a randomly distributed term to account for the heterogeneity across observations; all the $\beta_k$ coefficients are set up with non-informative priors as following normal distributions (0, 1E-6), and all the $\sigma_k^2$ are specified to be inverse-gamma priors as $\sigma_b^2 \sim \text{Inverse} - \text{gamma}(0.001, 0.001)$.

**Bayesian Random Parameters Conditional Logistic Model**

Suppose the crash occurrence has the outcomes $y_{ij}$=1 (crash event) and $y_{ij}$=0 (non-crash event) with respective probability $p_{ij}$ and 1-$p_{ij}$. The definitions of $i$ and $j$ are the same with Eq. (1). The random parameters conditional logistic regression can be expressed as follows:

$$y_{ij} \sim Bernoulli(p_{ij}) \tag{11}$$

$$logit(p_{ij}) = \alpha_i + \beta_{1i}X_{1ij} + \beta_{2i}X_{2ij} + \cdots + \beta_{Ki}X_{Kij} \tag{12}$$

$$\beta_{ki} = \beta_k + \varphi_{ki}, \qquad k = 0,1,2,\dots,K \tag{13}$$

$$\varphi_{ki} \sim N(0, \sigma_k^2) \tag{14}$$

Where $\alpha_i$ is the random intercept term for the ith stratum; $\boldsymbol{\beta_i} = (\beta_{1i}, \beta_{2i}, \dots, \beta_{Ki})$ is the vector of K random coefficients for the ith stratum; $\boldsymbol{X_{ij}} = (X_{1ij}, X_{2ij}, \dots, X_{Kij})$ is the vector of K independent variables for the jth observation in the ith stratum; $\varphi_{ki}$ is a randomly distributed term to account for the heterogeneity across strata; The main difference between random parameters logistic model and random parameters conditional logistic model is that the estimation of random parameters logistic model is based on classical likelihood function while random parameters conditional logistic model is based on the stratified conditional likelihood function (as shown in Eq. (4)). All the $\beta_k$ coefficients are also set up with non-informative priors as following normal distributions (0, 1E-6), and all the $\sigma_k^2$ are specified to be inverse-gamma priors as $\sigma_b^2 \sim \text{Inverse} - \text{gamma}(0.001, 0.001)$.

**Bayesian Inference and Model Comparisons**

Bayesian inference was employed in this study. For each model, three chains of 20,000 iterations were set up in WinBUGS (Lunn et al., 2000), the first 5,000 iterations were excluded as burn-in, the latter 15,000 stored iterations were set to estimate the posterior distribution. Convergence was evaluated using the built-in Brooks-Gelman-Rubin (BGR) diagnostic statistic (Brooks and Gelman, 1998).

The Deviance Information Criterion (DIC) can be used to compare complex models by offering a Bayesian measure of model fitting and complexity (Spiegelhalter et al., 2002). DIC is defined as outlined in Equation 15:

$$DIC = \overline{D(\theta)} + p_D \tag{15}$$

Where $D(\theta)$ is the Bayesian deviance of the estimated parameter, and $\overline{D(\theta)}$ is the posterior mean of $D(\theta)$. $\overline{D(\theta)}$ can be viewed as a measure of model fit, while $p_D$ denotes the effective number of parameters and indicates the complexity of the models. Models with smaller DIC are preferred. Very roughly, difference of more than 10 might definitely rule out the model with the higher DIC (Spiegelhalter et al., 2003).

In terms of model goodness-of-fit, the AUC value which is the area under Receiver Operating Characteristic (ROC) curve was also adopted. The ROC curve illustrates the relationship between the true positive rate (sensitivity) and the false alarm rate (1–specificity) of model classification results based on a given threshold from 0 to 1. It is worth noting that the classification results of Bayesian random parameters logistic model is based on the predicted crash probabilities, which lie in the range of 0 to 1, while the classification result of Bayesian conditional logistic model and Bayesian random parameters conditional logistic model are based on the predicted odds ratio, which may be larger than 1. In order to be consistent with the other two models, all the odds ratios predicted by Bayesian conditional logistic model were divided by the maximum odds ratio to create adjusted odds ratios. Later on, the adjusted odds ratios were used to create the classification result based on different threshold from 0 to 1. In this study, AUC values were calculated using R package pROC (Robin et al., 2011).

## MODELING RESULTS

This section discusses the modeling results of the Bayesian conditional logistic models based on four time slices datasets, followed by the model comparisons between Bayesian conditional logistic model, Bayesian random parameters logistic model, and Bayesian random parameters conditional logistic model based on the same dataset.

Four models based on 4 time-slice datasets are presented in TABLE 2. The model comparison results based on training and validation AUC values indicate that the slice 2 model (5-10 minute interval) performs the best, followed by the slice 1 (0-5 minute interval) model. However, based on slice 1 model, there would be no any spare time to implement any proactive traffic management strategy to prevent the possible crash occurrence. Moreover, as Golob and Recker (Golob et al., 2004) mentioned that there may exist 2.5 min difference between the exact crash time and reported crash time, thus the slice 1 model was treated as a reference. On the other hand, slice 2 model performs the best in terms of the number of significant variables. Finally, the slice 2 model was selected to conduct further interpretation and model comparison.

**TABLE 2. Model Results of Bayesian Conditional Logistic Regression Models based on Different Time Slices**

| Parameter | Slice 1 | | Slice 2 | | Slice 3 | | Slice 4 | |
|---|---|---|---|---|---|---|---|---|
| | Mean (95% BCI) | Hazard Ratio | Mean (95% BCI) | Hazard Ratio | Mean (95% BCI) | Hazard Ratio | Mean (95% BCI) | Hazard Ratio |
| Avg_speed | **-0.049** **(-0.071, -0.029)** | 0.952 | **-0.025** **(-0.048, -0.004)** | 0.975 | - | - | - | - |
| Up_Vol_LT | **0.024** **(0.007, 0.044)** | 1.024 | **0.024** **(0.005, 0.044)** | 1.024 | **0.024** **(0.006, 0.045)** | 1.024 | **0.036** **(0.014, 0.06)** | 1.037 |
| Down_GreenRatio | - | - | **-0.042** **(-0.075, -0.011)** | 0.959 | - | - | - | - |
| Rainy | **0.551** **(0.02374, 1.065)\*** | 1.735 | **0.667** **(0.055, 1.274)** | 1.948 | **0.682** **(0.037, 1.322)** | 1.978 | **0.72** **(0.078, 1.341)** | 2.054 |
| Training AUC | 0.6150 | | 0.6210 | | 0.5451 | | 0.5507 | |
| Validation AUC | 0.6081 | | 0.6169 | | 0.5300 | | 0.5476 | |

*Note: Mean (95% BCI) values marked in bold are significant at the 0.05 level; Mean (95% BCI) values marked in bold and noted by \* are significant at the 0.1 level.*

Based on the estimation results in the slice 2 model, four variables were found to be significantly associated with the crash occurrence on urban arterials: (1) the negative coefficient (-0.025) of average speed indicates that higher average speed tends to decrease the crash risk, which is consistent with other studies (Abdel-Aty et al., 2012; Ahmed et al., 2012a, b; Ahmed and Abdel-Aty, 2012; Shi and Abdel-Aty, 2015; Xu et al., 2012; Yu et al., 2016). This could be explained as the traffic condition with higher average speed, which represents more smooth traffic flow, could have better safety performance. Similarly, congestion index was found to have positive effect on crash likelihood (Hossain and Muromachi, 2012; Shi and Abdel-Aty, 2015), which means that the congested traffic condition is expected to have higher crash risk. The odds ratio of 0.975 means that when other variables held constant, one-unit increase in the average speed would decrease the odds of crash occurrence by 2.5%; (2) the upstream left-turn volume from the intersecting road segment was found to be positively correlated with crash likelihood, which might be explained in that more vehicles from the intersecting road segment left turning into the subject segment may result in more lane change behavior, which may lead to more conflicts with through vehicles. The odds ratio of 1.024 indicates that one-unit increase in upstream left-turn volume would lead to an increase of 2.4% in the odds of crash occurrence; (3) downstream green ratio was found to have negative effect on crash risk, and the odds ratio of 0.959 indicates that one percentage increase in downstream green ratio would decrease the odds of crash occurrence by 4.1%; (4) rainy indicator has a positive effect, the odds ratio of 1.948 means that the odds of crash occurrence under rainy condition is 94.8% higher than normal conditions, which is in line with previous studies (Ahmed et al., 2012a).

Furthermore, both Bayesian random parameters logistic model and Bayesian random parameters conditional logistic model were developed based on time slice 2 dataset, the model comparison results are as shown in TABLE 4. In the Bayesian random parameters logistic model, the upstream left-turn volume and downstream green ratio are insignificant, and this model has the lowest AUC value and the highest DIC value among the three models, these indicate that without considering the stratified data structure of the matched case-control dataset may significantly deteriorate the model performance. In order to improve the model performance of the Bayesian random parameters conditional logistic model, 15 ($\sum_{i=0}^{3} \frac{4!}{i!(n-i)!}$) combinations of fixed and random variables were developed to compare the model results, TABLE 3 shows the model performance of the 15 random parameter combinations.

**TABLE 3. Model Performance of Different Random Parameter Combinations**

| Model Type | Fixed Variables | Training AUC | Validation AUC |
|---|---|---|---|
| 4 random variables | - | 0.6217 | 0.6196 |
| 3 random and 1 fixed variables | Rainy | 0.6211 | 0.6155 |
| | Down_GreenRatio | 0.6202 | 0.6126 |
| | Up_Vol_LT | 0.6216 | 0.6232 |
| | Avg_speed | 0.6208 | 0.6134 |
| 2 random and 2 fixed variables | Avg_speed & Rainy | 0.6206 | 0.614 |
| | Avg_speed & Up_Vol_LT | 0.6208 | 0.6246 |
| | Avg_speed & Down_GreenRatio | 0.6209 | 0.6163 |
| | Up_Vol_LT & Down_GreenRatio | 0.622 | 0.6232 |
| | Up_Vol_LT & Rainy | 0.6215 | 0.6163 |
| | Down_GreenRatio & Rainy | 0.6208 | 0.6157 |
| 1 random and 3 fixed variables | Avg_speed & Up_Vol_LT & Down_GreenRatio | 0.6213 | 0.6164 |
| | Avg_speed & Up_Vol_LT & Rainy | 0.6216 | 0.6164 |
| | Avg_speed & Down_GreenRatio & Rainy | 0.6202 | 0.6119 |
| | Up_Vol_LT & Down_GreenRatio & Rainy | 0.6207 | 0.6158 |

Since all the modeling results of these 15 combinations will be too much to present, only the best model (i.e. fix "Up_Vol_LT" and "Down_GreenRatio", and randomize the other two variables) was presented in TABLE 4. Both the AUC and DIC values indicate that the Bayesian random parameters conditional logistic model performs better than the Bayesian conditional logistic model, which verified that introducing random parameters could improve model performance.

**TABLE 4. Model Comparison Results based on Time Slice 2**

| Parameter | Bayesian conditional logistic regression | | Bayesian random parameters logistic model | | Bayesian random parameters conditional logistic model | |
|---|---|---|---|---|---|---|
| | Mean (95% BCI) | Hazard Ratio | Mean (95% BCI) | Hazard Ratio | Mean (95% BCI) | Hazard Ratio |
| Intercept | - | - | **-1.514 (-2.35, -0.607)** | - | - | - |
| *Standard deviation* | - | - | *0.074 (0.021, 0.19)* | - | - | - |
| Avg_speed | **-0.025 (-0.048, -0.004)** | 0.975 | **-0.023 (-0.041, -0.006)** | 0.977 | **-0.027 (-0.051, -0.006)** | 0.973 |
| *Standard deviation* | - | - | *0.012 (0.009, 0.017)* | - | *0.044 (0.018, 0.091)* | |
| Up_Vol_LT | **0.024 (0.005, 0.044)** | 1.024 | 0.009 (-0.002, 0.021) | 1.01 | **0.025 (0.004, 0.047)** | 1.025 |
| *Standard deviation* | - | - | *0.017 (0.012, 0.024)* | - | - | - |
| Down_GreenRatio | **-0.042 (-0.075, -0.011)** | 0.959 | -0.007 (-0.017, 0.003) | 0.993 | **-0.045 (-0.076, -0.013)** | 0.956 |
| *Standard deviation* | - | - | *0.009 (0.007, 0.011)* | - | - | - |
| Rainy | **0.667 (0.055, 1.274)** | 1.948 | **0.797 (0.102, 1.436)** | 2.219 | **0.591 (0.082, 1.224)*** | 1.806 |
| *Standard deviation* | - | - | *0.070 (0.021, 0.17)* | - | *0.283 (0, 0.543)* | - |
| DIC | 682.290 | | 1179.610 | | 676.674 | |
| Training AUC | 0.6210 | | 0.5748 | | 0.6220 | |
| Validation AUC | 0.6169 | | 0.5714 | | 0.6232 | |

*Note: Mean (95% BCI) values marked in bold are significant at the 0.05 level; Mean (95% BCI) values marked in bold and noted by * are significant at the 0.1 level; The value in italic are the standard deviation of the corresponding parameter distribution.*

## CONCLUSION AND DISCUSSION

This study investigated the crash risk on urban arterials based on real-time data from multiple sources, including travel speed provided by Bluetooth detectors, traffic volume and signal phasing extracted from adaptive signal controllers, and weather data collected by the airport weather station. Matched case-control design with a control-to-case ratio of 4:1 was applied to collect data for crash and non-crash events. Four Bayesian conditional logistic models were developed separately for four 5-minute interval datasets (20-minute window prior to the reported crash time). In terms of AUC values, the model estimation results indicated that slice 2 (5-10 minute) model performs the best, followed by the slice 1 (0-5 minute) model. Considering that the implementation of proactive traffic management strategy may need some time in advance to possible crash occurrence, and there may exists error between the reported and actual crash times (Golob et al., 2004), slice 1 model was disregarded and slice 2 model was selected to conduct further analysis.

The results of the slice 2 model indicate that the average speed, upstream left-turn volume, downstream green ratio, and rainy indicator are significantly associated with the crash risk on urban arterials. In general, these finding are consistent with previous studies, in which the average speed was found to have significant negative impact on crash occurrence (Abdel-Aty et al., 2012; Ahmed et al., 2012a, b; Ahmed and Abdel-Aty, 2012; Shi and Abdel-Aty, 2015; Xu et al., 2012; Yu et al., 2016), while adverse weather (Ahmed et al., 2012a; Xu et al., 2013a) were found to be positively correlated with crash likelihood. In terms of the effect of traffic volume, only the upstream left-turn volume was found to have significant effect on crash likelihood, which indicates that more vehicles from the intersecting road segment left turning into the subject segment may increase the crash risk on the segment. This is quite different from the findings on freeways, which showed that the total upstream volume has significant positive impact on crash occurrence (Shi and Abdel-Aty, 2015; Yu et al., 2017; Yu et al., 2016).

It is worth noting that the downstream green ratio was found to be negatively associated with crash occurrence, this could be explained as the higher downstream green ratio could efficiently reduce the percentage of stop-and-go traffic, which may increase the safety performance. Surprisingly, the speed standard deviation is insignificant, this could be explained in that the average number of vehicles detected by the Bluetooth detector within 5-minute interval is about 6, which might be too small to capture the variation in speed.

Compared with the previous research on the real-time safety analysis of urban arterials (Theofilatos, 2017), they found that the 1 hour variation in both occupancy and volume were significantly associated with crash likelihood, which is quite different from our study. This might be explained in that the 1 hour aggregated traffic parameters can hardly represent the actual short-term traffic status such as speed and volume prior to crash occurrence, while it can capture the variation in traffic flow. This comparison implies that the traffic parameters should be aggregated based on more appropriate time interval, which can not only represent the short-term traffic status but also capture the variation in traffic flow characteristics.

Furthermore, the Bayesian random parameters logistic and Bayesian random parameters conditional logistic models were developed and compared with the Bayesian conditional logistic model based on the time slice 2 dataset. The results indicate that the Bayesian random parameters logistic model which ignored the stratified structure of the matched-case-control dataset performs the worst, which verifies that the stratified structure of the matched-case-control dataset should be considered in the modeling process. Moreover, the Bayesian random parameters conditional logistic model performs better than the Bayesian conditional logistic model, which demonstrates the advantage of random parameters model.

From the application point of view, the outcome of this study could be implemented from several aspects. The most straightforward application is to apply this algorithm to develop an arterial real-time crash risk prediction system. The real-time prediction results could be fed into the implementation of proactive traffic management strategies (e.g., variable speed limit), which can efficiently mitigate the crash risk in advance of the potential crash occurrence. Also, the real-time prediction results could be provided to drivers to assist with the route choice decisions. Furthermore, the real-time crash prediction results could be delivered to the drivers through connected-vehicle technology to provide crash risk warning information. In addition, the arterial real-time crash risk prediction system could be integrated with the real-time crash prediction on freeways. Therefore, an integrated arterial/freeway active traffic management strategy could be employed to proactively mitigate the safety of the road network.

However, the validation AUC value of 0.6232 implies that the model is still not ready to be applied to the real-time crash risk prediction and active traffic management system. In the future, more advanced machine learning techniques should be applied to improve the predictive performance. Nevertheless, the current estimation results could provide profound insights for traffic engineers to understand the relationship between crash risk and real-time traffic characteristics and weather conditions on arterials.

As this is the first attempt to investigate the real-time crash risk on urban arterials based on 5-minute aggregated data, there are still plenty of room for further improvement: (1) in order to achieve more accurate vehicle-level travel time and speed, the vehicle delay at intersections should be excluded from the travel time. In this context, high-resolution vehicle trajectory data would be preferable rather than Bluetooth data. (2) The current study focused on the safety effect of the traffic and signal status during different 5-minute intervals prior to the crash occurrence. Therefore, the exact signal status at the time of crash occurrence has not been considered. More disaggregate analyses, e.g., 1-min interval, should be conducted when higher resolution data are available. (3) As the Bluetooth data only provide the speed of the segment, it cannot distinguish the lane specific travel speed. In the future, lateral speed difference should be considered when more microscopic data are available. (4) This study only focused on the total crashes. Different crash types and crash severity could be considered in the future.